\documentclass[10pt,twocolumn,letterpaper]{article}

\usepackage{cvpr}              

\usepackage{algorithm}
\usepackage{algorithmic}

\usepackage{bm}
\usepackage{amsmath}
\usepackage{mathrsfs}
\usepackage{array}
\usepackage{algorithm}
\usepackage{booktabs}
\usepackage{array}
\usepackage{multirow}
\usepackage{multicol}
\usepackage{amssymb}

\definecolor{cvprblue}{rgb}{0.21,0.49,0.74}
\usepackage[pagebackref,breaklinks,colorlinks,allcolors=cvprblue]{hyperref}

\def\paperID{3430} 
\def\confName{CVPR}
\def\confYear{2025}

\title{Beyond Feature Mapping GAP: Integrating Real HDRTV Priors for Superior SDRTV-to-HDRTV Conversion}

\author{
 Kepeng Xu\textsuperscript{\rm 1\rm} ,
 Gang He\textsuperscript{\rm 1 \dag} ,
 Li Xu\textsuperscript{\rm 1 \dag},
 Siqi Wang\textsuperscript{\rm 1}\\
 Wenxin Yu\textsuperscript{\rm 2},
 Xianyun Wu\textsuperscript{\rm 1}\\
 \textsuperscript{\rm 1}Xidian University \\
 \textsuperscript{\rm 2}Southwest University of Science and Technology 
}

\begin{document}
\maketitle

\renewcommand{\thefootnote}{\fnsymbol{footnote}}
\footnotetext{$^\dag$Corresponding author.}
\footnotetext{email: kepengxu11@gmail.com}
\renewcommand{\thefootnote}{1}

\begin{abstract}
The rise of HDR-WCG display devices has highlighted the need to convert SDRTV to HDRTV, as most video sources are still in SDR. Existing methods primarily focus on designing neural networks to learn a single-style mapping from SDRTV to HDRTV. However, the limited information in SDRTV and the diversity of styles in real-world conversions render this process an ill-posed problem, thereby constraining the performance and generalization of these methods. Inspired by generative approaches, we propose a novel method for SDRTV to HDRTV conversion guided by real HDRTV priors. Despite the limited information in SDRTV, introducing real HDRTV as reference priors significantly constrains the solution space of the originally high-dimensional ill-posed problem. This shift transforms the task from solving an unreferenced prediction problem to making a referenced selection, thereby markedly enhancing the accuracy and reliability of the conversion process. Specifically, our approach comprises two stages: the first stage employs a Vector Quantized Generative Adversarial Network to capture HDRTV priors, while the second stage matches these priors to the input SDRTV content to recover realistic HDRTV outputs. We evaluate our method on public datasets, demonstrating its effectiveness with significant improvements in both objective and subjective metrics across real and synthetic datasets.
\end{abstract}

\begin{figure*}[htbp]
    \centering
    \includegraphics[width=0.999\textwidth]{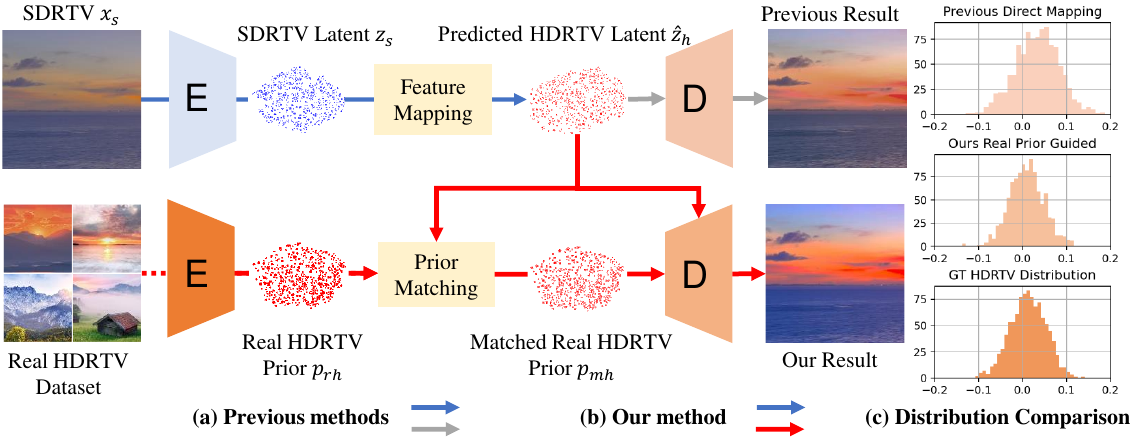}
    \caption{
(a) Previous methods learn single-style SDRTV-to-HDRTV conversion on a single dataset. However, the SDRTV-to-HDRTV conversion distribution in real-world scenarios is complex and diverse, which makes it difficult for previous methods to effectively convert SDRTV-to-HDRTV conversion in the real-world.
(b) Our method embeds rich and realistic HDRTV into the converted neural network, thereby greatly improving the conversion performance in real scenes.
(c) The latent variable distribution of our method is closer to GT due to the incorporation of real HDRTV prior guidance. 
    }
    \label{fig1comparewithpreintro}
\end{figure*}

\section{Introduction}

The dynamic range of a video, defined by the difference between its maximum and minimum luminance, enables High Dynamic Range (HDRTV) to deliver superior visuals. Advances in Electro-Optical Transfer Functions (EOTF), such as PQ/HLG, and Wide Color Gamut (WCG) RGB primaries (as per BT.2020), enhance HDR technology's potential.

Despite the rise of WCG-HDR displays, the production complexities result in limited WCG-HDR content. Consequently, many films remain in Standard Dynamic Range (SDRTV), driving the demand for SDRTV-to-HDRTV conversions. HDRTV offers a wider color gamut (Rec. 2020 vs Rec. 709), higher brightness range (0.01-1000 nits vs 0.1-100 nits), advanced EOTF curves (PQ/HLG vs Gamma), and greater color depth (>10-bit vs 8-bit). However, the scarcity of HDRTV content compared to SDRTV makes SDRTV-to-HDRTV conversion essential, despite the inherent challenges due to the limitations of existing imaging systems and transmission protocols.

Traditional methods \cite{Huo2013PhysiologicalIT, Kovaleski2014HighQualityRT} suffer from color inaccuracies and abnormal brightness restoration due to limitations in estimating curve parameters for SDRTV to HDRTV conversion. Meanwhile, recent neural network-based approaches \cite{kim2019deep, kim2020jsigan, sritmgan, chen2021hdrtvnet, kpnmfi, fmnet6, hyconditm} employ the strategy of encoding SDRTV content into a latent space and subsequently reconstructing it as HDRTV content. 
Models designed by these previous methods are trained and tested on a single data set, as shown in Figure \ref{fig1comparewithpreintro} (a). 

However, SDRTV to HDRTV conversion models trained on a single dataset are difficult to adapt to the content diversity of the real world. This challenge arises because these models learn a fixed mapping that is inherently tied to the specific characteristics of the dataset on which they are trained.
Actually, these characteristics can include, but are not limited to, lighting conditions, types of scenes, and tone mapping schemes.

\textbf{Insights.}
In the process of video production, a single HDRTV content might correlate with various SDRTV-style versions. This scenario underscores the complexity faced when training neural networks to understand not just a single linear relationship. 
Due to the ill-posed of multiple mapping relationships and the lack of deterministic mapping rules, it is difficult for neural networks to learn such chaotic mapping functions.
This complexity highlights the substantial challenge in developing neural networks that can directly learn the diverse SDRTV-to-HDRTV conversions reflective of real-world scenarios.
Therefore, it is very difficult to build a neural network to directly learn real-world SDRTV-to-HDRTV conversion.

\textbf{Previous Solution.}
HDRTVDM\cite{Guo_2023_CVPR} emphasizes the importance of dataset diversity for training neural networks that more closely match real-world content. Although the complexity of these datasets is close to actual conditions, existing neural networks face difficulties in fully capturing and learning the diverse mapping relationships present within the dataset. This situation calls for improvements in neural network architectures to better process and understand the rich, complex data reflective of real-world scenarios.

\textbf{Our Solution.}
In contrast, our proposed RealHDRTVNet framework enhances the quality of SDRTV to HDRTV conversion by directly embedding HDRTV priors into the transformation process, as illustrated in Fig.\ref{fig1comparewithpreintro} (b). This approach effectively transforms the ill-posed restoration problem into a prior selection problem, significantly reducing the solution space size. By leveraging rich and diverse HDRTV priors, our method overcomes previous limitations, achieving more accurate, generalized, and reliable SDRTV to HDRTV mapping.

Moreover, a significant challenge in SDRTV-to-HDRTV conversion is the assessment of the perceptual quality of neural network-generated HDRTV content. 
Common metrics like LPIPS \cite{Zhang2018LPIPS}, NIQE \cite{niqeref}, and FID \cite{heusel2017gans}, which are typically used for SDRTV quality evaluation,  struggle to capture HDRTV's unique features within the PQ EOTF curve and Rec.2020 color gamut.

Inspired by this, our work extend tailored metrics for precise HDRTV quality assessment, including Learned Perceptual HDRTV Patch Similarity (LPHPS), Natural HDRTV Quality Evaluator (NHQE), and Fréchet Initial Distance (FHAD). These metrics are specifically designed to evaluate the subjective quality of HDRTV content directly.
With these innovative metrics, both researchers and practitioners have the tools to conduct reliable subjective quality evaluations of HDRTV content.

This paper's contributions are in three parts:

\begin{itemize}

\item {We propose an a priori selected SDRTV to HDRTV conversion method, which significantly limits the solution space of the original high-dimensional ill-posed problem, thereby enabling efficient learning of real-world SDRTV to HDRTV conversion and improving the quality of the converted HDRTV.}

\item {We quantitatively and qualitatively demonstrate that our proposed method outperforms previous methods.}
\end{itemize}

\begin{figure*}[htbp]
    \centering
    \includegraphics[width=0.999\textwidth]{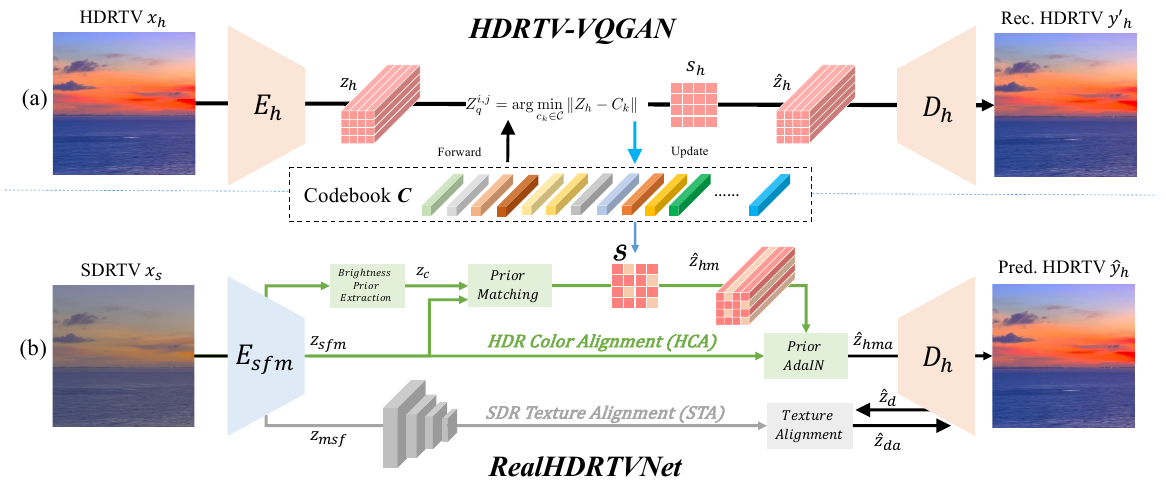}
    \caption{RealHDRTVNet framework. 
    (a) HDRTV-VQGAN. We first pre-train an HDRTV-VQGAN to learn to store HDRTV priors through self-reconstruction.
    (b)RealHDRTVNet. The learning modulation encoder $E_{sfm}$ obtains ``nearly high-quality HDRTV features''. Next, the HDR Color Alignment HCA module aligns the input features with HDRTV in the color dynamic range dimension. In addition, the SDR Texture Alignment STA module is used to align the texture with the input SDRTV. This makes the dynamic range information of the conversion result consistent with HDRTV, and the texture details consistent with SDRTV.}
    \label{fig1comparewithpre}
\end{figure*}

\section{Related work}

\subsection{SDRTV-to-HDRTV Methods}

\subsubsection{Our Research Scope.}

Recent advancements in HDR imaging have seen the advent of various learning-based methods, as noted by Wang and Yoon \cite{Wang_Yoon_2022}. These methods have found different applications in the real world. HDR enhancement primarily involves the use of neural networks to convert SDR images to HDR \cite{Kovaleski2014HighQualityRT}. On the other hand, Multi-Exposure HDR Imaging employs exposure bracketing to create HDR images from a sequence of SDR images taken at different exposure levels \cite{Chaudhari2019MergingISP,Le2022SingleImageHDR,Xu2021MultiExposureFusion,Chen2021AttentionGuidedHDR}. This paper concentrates on the transformation of SDRTV to HDRTV, aiming to achieve an enhanced dynamic range and wide color gamut in videos.

\subsubsection{DNN-based SDRTV-to-HDRTV Methods.}
Recent studies primarily focus on devising feature modulation strategies for robust SDRTV-to-HDRTV conversion \cite{kim2019deep,kim2020jsigan,sritmgan,chen2021hdrtvnet,kpnmfi,fmnet6,hyconditm}. These strategies entail the utilization of diverse color-prior techniques to facilitate effective feature modulation. The HDRTVDM approach \cite{Guo_2023_CVPR} notably enhances HDRTV conversion quality by refining the dataset.
In terms of integrated frameworks, SR-ITM \cite{kim2019deep} presents a network design that concurrently addresses SDRTV-to-HDRTV transformation and super-resolution. Similarly, DIDnet \cite{xu2023towards} introduces a network model that combines SDRTV-to-HDRTV conversion with the restoration of coding artifacts.

\label{datasetlabel}

\subsubsection{SDRTV-to-HDRTV Datasets.}
Within the current research landscape, a mere quartet of open-source HDRTV datasets\cite{kim2019deep,chen2021hdrtvnet,Guo_2023_CVPR} exists. Each dataset adheres to the BT.2020 RGB color gamut. Further, they conform to the PQ EOTF specifications with a luminance zenith of 1000 nits.

\subsection{Generative Adversarial Network}
The introduction of Generative Adversarial Networks (GANs), as demonstrated in Goodfellow et al.'s 2014 work \cite{goodfellow2014generative}, has revolutionized tasks like image restoration \cite{wang2021towards}. GAN priors, known for capturing complex image distributions, contrast traditional methods by integrating natural image characteristics into degraded visuals through adversarial training. This approach has advanced subfields like denoising, inpainting, super-resolution \cite{wang2018esrgan}, and artifact reduction \cite{wan2020bringing}. The strength of GAN priors lies in their restoration ability and in producing outputs closely aligned with original images. Additionally, models like VQGAN have been extensively applied in areas like multi-modal conversion \cite{Esser2021TamingTF}.

\section{Methodology}

\subsection{Preliminary} 
The conversion from SDRTV to HDRTV can be mathematically represented by the Maximum A Posteriori (MAP) estimate $p(h\mid s)$, which relies on the distribution of the SDRTV input $p(s)$. Traditional methods typically utilize inverse tone mapping functions to achieve this conversion. This process is formally defined in Equation \ref{invf}:

\begin{equation}
    \hat{h} = f(s; \theta_c) + \epsilon,
    \label{invf}
\end{equation}

\noindent
where $\hat{h}$ represents the reconstructed HDRTV, $\theta_c$ denotes the estimated curve parameters, and $\epsilon$ accounts for the error due to parameter estimation inaccuracies. This method often leads to color distortions and anomalies in brightness restoration in the converted HDRTV content.

Modern neural network-based methods \cite{kim2019deep,kim2020jsigan,sritmgan,chen2021hdrtvnet,kpnmfi,fmnet6,hyconditm} utilize a three-stage process to complete SDRTV-to-HDRTV conversion. Initially, an encoder module is employed to map the SDRTV content into a latent representation. Subsequently, the SDRTV latent features are input to the feature transformation module to obtain HDRTV latent features. The final stage involves a decoder module, which reconstructs the latent representation back into HDRTV. 
By minimizing the difference with real samples, the distribution of reconstructed HDRTV is close to the distribution of real data.
This dependency can be modeled as a conditional probability distribution $P(h|s)$, which is only influenced by the distribution of the input SDRTV $P(s)$:

\begin{equation}
p_\theta(h|s) =\int p_{\theta_d}(h|\hat{z_h})p_{\theta_{\tau}}(\hat{z_h}|z_s)p_{\theta_{e}}(z_s|s)ds \\
\end{equation}

\noindent

where:
\begin{itemize}
  \item \( \theta_e \), \( \theta_{\tau} \), and \( \theta_d \) represent the parameters of the learned encoder module, feature transformation module, and decoder module of the neural network.
  \item \( z_s \) is the initial latent representation into which the SDRTV image \( s \) is encoded.
  \item \( \hat{z_h} \) is the intermediate latent representation obtained after encoding and feature transformation of the SDRTV image \( s \).

\end{itemize}

These methods aim to learn a straightforward and static mapping relationship from SDRTV to HDRTV, inadequately capturing the intricate and multifaceted conversion process between SDRTV and HDRTV.
HDRTVDM\cite{Guo_2023_CVPR} believes that increasing the complexity of the dataset can force the neural network to learn an SDRTV-to-HDRTV mapping function that is more in line with the real world. Although increasing the dataset is closer to the real world at the sample level, it is difficult for the neural network to directly learn such a complex mapping process.

Transitioning from SDRTV to HDRTV inherently involves conditional probability transition. On a singular dataset, these conditional transitions adhere to a stationary distribution, making it feasible for neural networks to learn such probabilistic mappings with relative ease. However, in the real-world scenario, the SDRTV-to-HDRTV conditional probability transitions do not conform to a single, fixed distribution, posing a challenge for neural architectures to learn this intricate probability transition.

To address this challenge, 
we propose a framework that can directly harness the prior knowledge encapsulated within HDRTV data to facilitate a superior-quality transformation from SDRTV to HDRTV. 
The conversion can be described by the following equation.

\begin{equation}
\begin{split}
p_\theta(h|s,\pi) = \int & p_{\theta_d}(h|z_m)p_{\theta_m}(z_m|\hat{z_h},\pi) \\
&  p_{\theta_{\tau}}(\hat{z_h}|z_s)p_{\theta_{e}}(z_s|s) \, ds 
\end{split}
\end{equation}
\noindent
where:
\begin{itemize}
  \item \( \theta \) is the set of network parameters.
  \item \( z_m \) is the best matching HDRTV prior representation selected from the HDRTV prior distribution \( \pi \).
  \item \( \hat{z_h} \) is the intermediate latent representation obtained after encoding and mapping the SDRTV image \( s \).
  \item \( \pi \) is the HDRTV prior set .
  \item \( z_s \) is the initial latent representation into which the SDRTV image \( s \) is encoded.
\end{itemize}

To circumvent the traditional reliance solely on SDRTV inputs, our approach ingeniously integrates HDRTV priors into the conversion workflow, thereby facilitating efficient and complex HDRTV reconstruction. Our method enables adaptive conditional probability transitions based on the embedded HDRTV priors, moving beyond fixed-type conditional transitions to achieve high-quality conversion from SDRTV inputs.

\subsection{Overall Framework}

Following the instantiation of our motivation, we propose RealHDRTVNet, a novel architecture designed to learn the complex and variable transformations from SDRTV to HDRTV, demonstrating superior generalization capabilities in authentic scenarios.
Our methodology unfolds in three phases: initially, HDRTV-VQGAN is trained to embed real HDRTV priors. The subsequent two phase focuses on the SDRTV-to-HDRTV transformation, leveraging the pre-embedded HDRTV priors to augment the quality of HDRTV.
The embedded HDRTV prior can serve as a powerful guide to ensure high-quality HDRTV restoration.
The proposed method no longer learns a simple function mapping that only relies on SDRTV, but learns a transformation process guided by a real HDRTV prior. Guided by real HDRTV, our method can learn complex and diverse SDRTV-to-HDRTV conversion relationships.
Based on this, our method adopts a three-phase strategy:

\begin{itemize}
    \item {{Phase I}: We train a VQGAN model on HDRTV domain, embedding real HDRTV priors.}

    \item {{Phase II}: We craft a preliminary modulation encoder for SDRTV to HDRTV transformation. Through feature modulation techniques, this model refines the SDRTV latent distribution towards the anticipated HDRTV latent space, ensuring that the next stage can more accurately match the HDRTV prior.}

    \item {{Stage III}: We propose the HDR Color Alignment module HCA and the SDR Texture Alignment module STA. HCA identifies the best HDRTV prior from the pre-trained VQGAN codebook and uses the identified HDRTV to assist the conversion process. The SDR Texture Alignment module STA aligns the transformed features with SDRTV in texture to ensure texture fidelity.}
\end{itemize}

With this three-stage approach, we provide a novel solution for high-quality SDRTV-to-HDRTV conversion.



\subsection{Phase I: HDRTV Vector Quantized AutoEncoder - HDR Prior Representation Learning}
To reduce the uncertainty of SDRTV-to-HDRTV mapping and complement high-quality HDRTV color information, we first train a vector quantized autoencoder to learn a context-rich codebook that improves the quality of the converted HDRTV.

The specific structure is shown in Fig. \ref{fig1comparewithpre} (a). First, the HDRTV SDRTV $x_h \in R^{H \times W \times 3}$ is input into the encoder $E_h$ to get the latent feature $z_h \in R^{h \times w \times c}$.
Next, find the closest codebook feature $\hat{z}_h$ to $z_h$ in codebook $C \in R^{n \times k}$ (n is the number of vectors in the codebook) and the corresponding codebook index $S$.  Then $\hat{z}_h$ is fed into the decoder $D_h$ to obtain the reconstructed HDRTV $y'_{h}$.

We describe the details of  HDRTV-VQGAN.
First use the encoder $E_h$ to encode the input HDRTV $x_h$ into a latent representation $z_h$.
\begin{equation}
    z_h = E_h(x_h), 
\end{equation}
\noindent
where \( x_h \in \mathbb{R}^{H \times W \times 3} \) represents the input HDRTV, and \( z_h \in \mathbb{R}^{h \times w \times c} \) denotes the latent feature obtained from the encoder \( E_h \).

Next, the replaced features $\hat{z}_h$ and corresponding index $s_h$ are obtained from the codebook $C$ through nearest neighbor matching;
\begin{equation}
    \hat{z}_h, s_h = \underset{c_i \in C}{\mathrm{arg min}} \, \| z_h - c_i \|,
\end{equation}
\noindent
where $C \in \mathbb{R}^{n \times d}$ is the codebook containing $n$ vectors, each of dimension $d$, and $\hat{z}_h$ is the codebook feature closest to $z_h$. The corresponding codebook index is obtained as $S = \text{index of } \hat{z}_h$.

Then,$\hat{z}_h$ is processed via the decoder $D_h$  get converted HDRTV:
\begin{equation}
    y'_{h} = D_h(\hat{z}_h),
\end{equation}

\subsection{Phase II: SDRTV Modulation Encoder - Preliminary HDR Mapping}

Given the substantial differences in dynamic range and color space between SDRTV and HDRTV, direct alignment of SDRTV features to HDRTV priors in latent space is inherently challenging. To mitigate this, we decompose the prior matching problem into a two-stage process.

First, we propose the SDR Feature Modulation Encoder ($E_{sfm}$) to transform SDRTV latent features into a space that is more congruent with HDRTV priors. The encoder $E_{sfm}$ is composed of four sequential SDRTV Feature Modulation (SFM) blocks, which iteratively refine the latent representation $z_{sfm}$ to approximate the HDRTV distribution. The detail of the architecture is illustrated in the appendix and consists of residual blocks, SFM blocks, and downsampling layers.

Formally, the input SDRTV frame $x_s$ (referred to as  $x_0$ here) is processed as follows:

\begin{equation}
x_i = \text{Down}(\text{SFM}(\text{ResBlock}_c(x_{i-1}))),
\end{equation}
where $i\in \{1,2,3,4\}$, and $c \in \{64,128,256,512\}$, representing the number of channels at each respective stage.
In detail, the SFM is implemented by: 
\begin{equation}
\hat{x}_i = \alpha_i \odot x_i + \beta_i; \ \ \alpha_i, \beta_i = \text{Conv}_\theta(x_i),
\end{equation}
where $\odot$ denotes element-wise multiplication, $\alpha_i$ and $\beta_i$ denote the modulation parameters derived via convolutional layers $\text{Conv}_\theta$.

Subsequently, a Transformer module aggregates the modulated features into a compact latent representation:

\begin{equation}
    z_{sfm} = \text{ResBlock}_{c512}(\text{Transformer}(x_{m_4})).
\end{equation}

This process effectively aligns the SDRTV latent space with that of HDRTV, thereby facilitating optimal prior matching in subsequent processing stages.

\begin{figure*}[ht]
    \centering
    \includegraphics[width=0.99\textwidth]{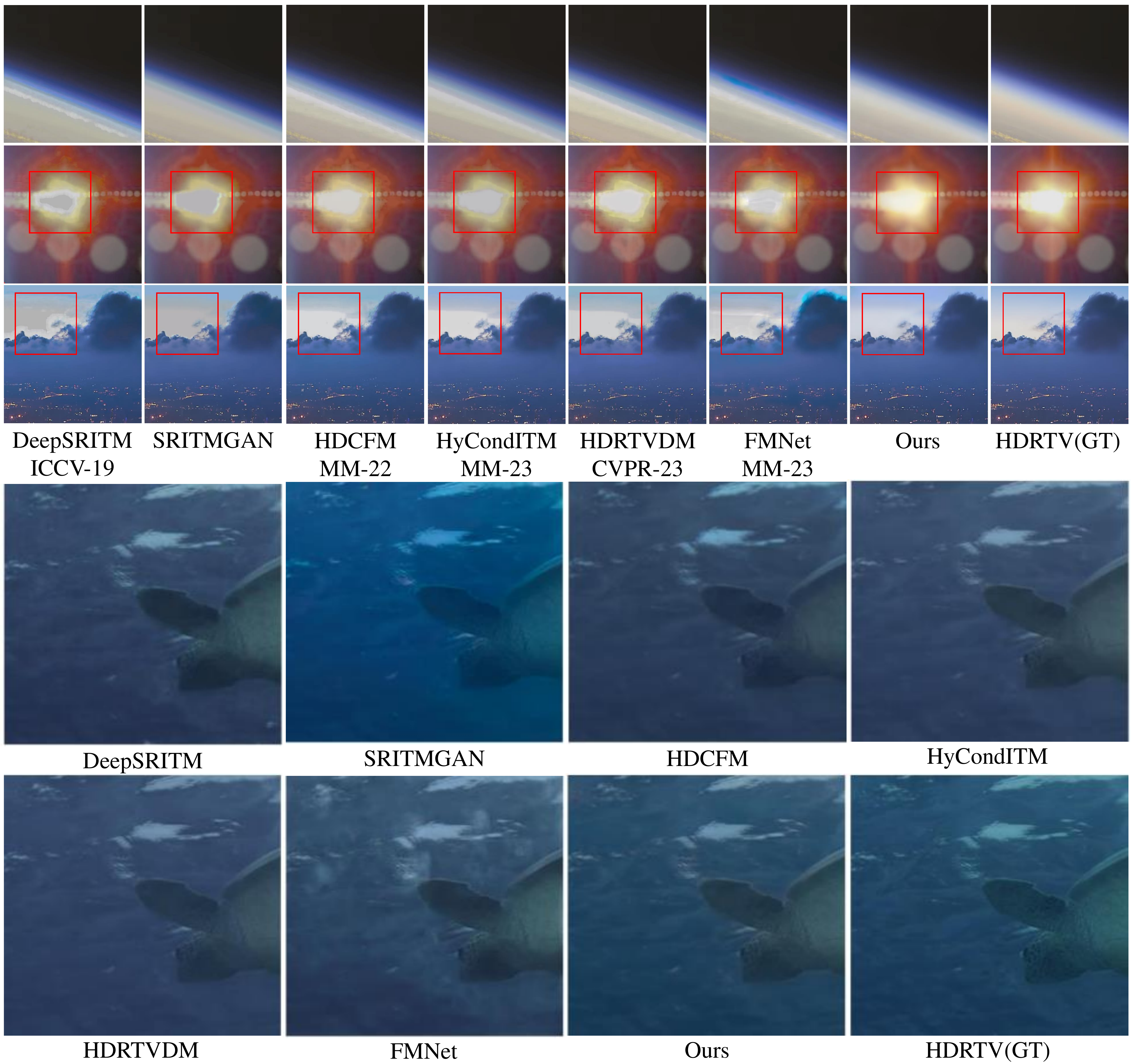}
    \caption{Qualitative results on synthetic datasets. Our RealHDRTVNet can recover realistic HDRTV color information through embedded real-world HDRTV priors. \textbf{(Zoom in for details)}}
    \label{figinsyn}
\end{figure*}

\begin{table*}[h]
  \centering
  \caption{Quantitative results on synthetic three SDRTV-to-HDRTV datasets.}
  \resizebox{\textwidth}{!}{%
    \begin{tabular}{>{\centering\arraybackslash}p{1.5cm}
    >{\centering\arraybackslash}p{0.8cm}
    >{\centering\arraybackslash}p{0.7cm}
    >{\centering\arraybackslash}p{0.7cm}
    >{\centering\arraybackslash}p{0.7cm}
    >{\centering\arraybackslash}p{0.8cm}|
    >{\centering\arraybackslash}p{0.7cm}
    >{\centering\arraybackslash}p{0.7cm}
    >{\centering\arraybackslash}p{0.7cm}
    >{\centering\arraybackslash}p{0.7cm}
    >{\centering\arraybackslash}p{0.8cm}|
    >{\centering\arraybackslash}p{0.7cm}
    >{\centering\arraybackslash}p{0.7cm}
    >{\centering\arraybackslash}p{0.7cm}
    >{\centering\arraybackslash}p{0.7cm}
    >{\centering\arraybackslash}p{0.8cm}
    }
    \toprule
    \multirow{2}{*}{Methods}&\multicolumn{5}{c|}{Test on HDRTV1K Dataset}   & \multicolumn{5}{c|}{Test on HDRTV4K Dataset} & \multicolumn{5}{c}{Test on SRITM Dataset} \\
         & PSNR  & SSIM  & LPHPS & NHQE  & FHAD  & PSNR  & SSIM  & LPHPS & NHQE  & FHAD  & PSNR  & SSIM  & LPHPS & NHQE  & FHAD \\
    \midrule
    IRSDE & 22.64 & 0.8683 & 0.3189 & 4.35  & 121.41 & 16.71 & 0.7663 & 0.3321 &  {4.45} & 119.10 & 22.07 & 0.8721 & 0.2789 & 4.53  & 113.31 \\
    DeepSRITM & 29.79 & 0.8945 & 0.3260 & 4.09  & 131.56 & 23.61 & 0.6925 & 0.4373 &  {4.42} & 143.58 & 27.8  & 0.8767 & 0.3423 & 4.28  & 129.34 \\
    FMNet & 34.18 & 0.9527 & 0.1812 & 3.94  & 103.78 & 27.43 & 0.8414 & 0.3077 & 5.95  & 124.06 & 29.26 & 0.9042 & 0.221 & 3.99  & 96.27 \\
    HDCFM & 32.42 & 0.9414 & 0.1714 &  {3.92} & 104.82 & 25.41 & 0.8404 & 0.2541 & 4.49  & 112.20 & 28.44 & 0.871 & 0.2157 &  \textbf{3.83} & 96.22 \\
    HDRTVDM & 34.09 & 0.9268 & 0.2189 & 4.22  &  {95.73} & 27.47 &  {0.8792} &  {0.2286} & 5.38  &  {102.27} &  {31.07} &  {0.9138} &  {0.2102} & 4.14  &  {85.73} \\
    HDRTVNet &  \textbf{36.01} &  {0.9559} &  {0.1593} & 3.98  & 100.42 &  {28.06} & 0.8368 & 0.2836 & 5.64  & 121.20 & 29.47 & 0.8747 & 0.2327 &  {3.94} & 92.84 \\
    HyConDITM &  {35.61} &  {0.9566} &  {0.1288} &  {3.91} &  {95.33} &  {29.31} &  {0.8702} &  {0.2064} & 5.57  &  {102.61} &  \textbf{31.16} & {0.9176} &  {0.1555} &  {3.96} &  {84.24} \\
    Ours  &  {35.06} &  \textbf{0.9609} &  \textbf{0.1166} &  \textbf{3.88} &  \textbf{91.03} &  \textbf{30.31} &  \textbf{0.8912} &  \textbf{0.1804} &  \textbf{4.03} &  \textbf{96.15} &  {29.65} &   \textbf{0.9308} &  \textbf{0.1392} &  {3.94} &  \textbf{81.56} \\
    \bottomrule
    \end{tabular}%
    }  
    
  \label{tab:addlabel1}%
\end{table*}%

\subsection{Phase III: RealHDRTVNet - High-Quality Conversions with Pretrain HDR Prior}
\label{maskcalll}

In phase I and II, $E_{sfm}$, the codebook $C$, and $D_h$ are pre-trained and then frozen in the subsequent phase. 
In phase III, to achieve efficient and nuanced HDRTV reconstruction, we propose RealHDRTVNet, which consists of two key components: \textbf{HDR Color Alignment Module(\textit{HCA})} and \textbf{SDR Texture Alignment(\textit{STA})}. These components ensure that the converted HDRTV retains the texture structure of the input SDRTV while aligning with the dynamic range and color of real HDRTV.

\subsubsection{HDR Color Alignment Module(HCA).}
To achieve accurate color alignment with real HDRTV priors, we introduce the HDR Color Alignment module(HCA), which integrates the functionalities of brightness prior extraction, prior matching, and prior adaptive instance normalization.

Given an SDRTV input $x_s$, the encoder $E_{sfm}$ first extracts the multi-scale feature $z_{msf}$ and the “basic” HDRTV feature $z_{sfm}$. Recognizing the necessity for adaptive processing in highlight regions, we employ a brightness prior extraction module to generate a luminance-aware position coding $z_c$ using $X_s$ and its highlight mask. This feature $z_c$ is then fed into the Prior Matching (PM) module, where it is aligned with the optimal HDRTV prior $\hat{h}_{hm}$ from the codebook $C$, resulting in the matched feature $\hat{z}_{hm}$.

After obtaining $\hat{z}_{hm}$, Prior Adaptive Instance Normalization (AdaIN) adjusts the latent features of $z_{sfm}$ and $\hat{z}_{hm}$, generating $\hat{z}_{hma}$. The feature $\hat{z}_{hma}$ is then fed into the pretrained $D_h$ to produce the HDRTV result. This process ensures that the color and dynamic range of the generated HDRTV align with real HDRTV priors.

\subsubsection{SDR Texture Alignment Module(STA)}.
To maintain the texture structure of the input SDRTV in the converted HDRTV, we introduce the SDR Texture Alignment module. This module uses multi-scale features $z_{msf}$ from the encoder $E_{sfm}$ during the decoding process to align textures.

First, the multi-scale features $z_{msf}$ from the encoder are concatenated with the decoder features $\hat{z}_{d}$ and fed into a deformable convolution to achieve alignment, resulting in aligned features $\hat{z}_{n}$. These aligned features are then used to modulate the decoder features $\hat{z}_{d}$, resulting in the aligned decoder feature $\hat{z}_{da}$, which is fed into decoder $D_h$ for HDRTV reconstruction. 

By performing SDR Texture Alignment on the decoder at four different resolutions, the decoder $D_h$ is capable of recovering HDRTV with a realistic dynamic range while preserving high-quality texture details from the SDRTV.

\subsubsection{Overall Pipeline}
The complete RealHDRTVNet pipeline is described as follows:
\begin{align}
    z_{sfm}, z_{msf} & = E_{sfm}(x_s), \\
    \hat{z}_{hma} & = HCA(z_{sfm}, x_s, C), \\
    \hat{y}_h & = D_h(\hat{z}_{hma}, STA(z_{msf}, \hat{z}_d)).
\end{align}

\section{Experiment}

\subsection{Evaluation Metrics}
\label{sectsubmetric}

We use objective metrics, subjective metricsand user study to evaluate different methods.
Objective metrics include PSNR, SSIM, $\Delta E_{ITP}$ and HDRVDP3, which respectively evaluate the fidelity, structural similarity, color fidelity, and visual similarity of the converted HDRTV.
Subjective metrics include LPHPS(Extended from LPIPS), FHAD(Extended from FID), and NHQE(Extended from NIQE), which can evaluate the perceptual similarity and distribution consistency of the converted HDRTV and the real HDRTV.
These three subjective metrics are obtained by expanding the previous subjective quality assessment methods in this paper. The detailed process is shown in the appendix.

\subsection{Implementation Details}

We validate our method's efficacy by training and testing on various datasets detailed in Section \ref{datasetlabel}, and additionally evaluating real SDRTV datasets. Our approach employs the Adam optimizer with an initial learning rate of  $2 \times 10^{-5}$, utilizing a cosine annealing schedule for learning rate decay. Training is divided into three phases, which means HDRTV-VQGAN, $E_{sfm}$, and RealHDRTVNet are trained respectively. The details are given in the appendix.

\begin{table*}[htbp]
    \caption{Quantitative results of Non-Reference metrics (FHAD and NHQE) on five real-world SDR datasets: BSD100\cite{937655}, CBSD68-Noisy\cite{937655}, CBSD68-Original\cite{937655}, Set14\cite{set14bi}, and Set5\cite{set5bi}.}
  \centering
    \resizebox{\textwidth}{!}{
    \begin{tabular}{ccccccccccc}
      \toprule
      Datasets & \multicolumn{2}{c}{BSD100} & \multicolumn{2}{c}{CBSD68-Noisy} & \multicolumn{2}{c}{CBSD68-Original} & \multicolumn{2}{c}{Set14} & \multicolumn{2}{c}{Set5} \\
    \midrule
    Metrics & NHQE $\downarrow$  & FHAD  $\downarrow$  & NHQE  $\downarrow$  & FHAD  $\downarrow$  & NHQE  $\downarrow$  & FHAD  $\downarrow$  & NHQE  $\downarrow$  & FHAD  $\downarrow$  & NHQE  $\downarrow$  & FHAD  $\downarrow$  \\
    \midrule
    DEEPSRITM & 3.73  & 147.06 & 3.72  & 144.58 & 3.67  & 148.92 & 3.68  & 176.66 & 3.80  & 193.45 \\
    FMNET & 3.94  & 147.33 & 4.53  & 148.84 & 3.95  & 149.39 & 4.11  & 181.84 & 3.76  & 203.90 \\
    HDCFM & 3.81  & 145.67 & 3.73  & 147.43 & 3.80  & 146.81 & 3.70  & 175.99 & 3.79  & 198.89 \\
    HDRTVDM & 3.76  & 144.27 & 3.75  & 144.89 & 3.78  & 145.50 & 3.67  & 178.58 & 3.75  & 198.48 \\
    HDRTVNET & 3.84  & 150.68 & 3.82  & 146.08 & 3.82  & 151.66 & 3.72  & 177.90 & 3.82  & 199.08 \\
    HyCondITM & 3.74  & 141.63 & 3.68  & 142.72 & 3.73  & 144.49 & 3.66  & 174.74 & 3.85  & 198.41 \\
    Ours  &  \textbf{3.61} &  \textbf{141.43} &  \textbf{3.61} &  \textbf{142.40} &  \textbf{3.60} &  \textbf{144.08} &  \textbf{3.57} &  \textbf{174.41} &  \textbf{3.71} &  \textbf{192.61} \\
    \bottomrule
    \end{tabular}%
    }

  \label{tabrealtesttable}%

\end{table*}%

\begin{table}[t]
  \centering
  \caption{Results on HDRVDP3 metric.}
    \begin{tabular}{cccc}
    \toprule
          & HDRTV1K  & HDRTV4K  & SRITM  \\
    \midrule
    IRSDE & 6.45  & 5.60  & 6.58 \\
    HDCFM & 7.01  & 6.64   & 6.43 \\
    HDRTVDM   & 7.77   & 6.68  &  7.53 \\
    HDRTVNet  & 7.64  & 6.27   & 6.60 \\
    HyConDITM & 8.22 &7.58 & 7.71 \\
    Ours   & 8.28  & 7.91 & 7.80 \\
    \bottomrule
    \end{tabular}%
  \label{tabhdrvdp3}%
\end{table}%

\subsection{Comparisons with State-of-the-Art Methods} 
We compared our proposed RealHDRTVNet with state-of-the-art methods, including HDRTVNet\cite{chen2021hdrtvnet}, HyCondITM\cite{hyconditm}, HDCFM\cite{hdcfmcite}, HDRTVDM\cite{Guo_2023_CVPR}, FMNet\cite{fmnet6}, and DEEPSRITM\cite{kim2019deep}. Extensive experiments were conducted on both synthetic and real-world datasets.

\subsubsection{Experimental Results on Synthetic Datasets}
We first show quantitative results on three synthetic datasets in Table \ref{tab:addlabel1}. On quality metrics PSNR, SSIM, LPHPS, FHAD, and NHQE, our RealHDRTVNet achieves the best scores than existing methods. This means that our method achieves the best performance and produces HDRTV results of higher subjective quality.

Our method achieves a lower value on the LPHPS metric, indicating that the perceptual difference between the proposed method and the real HDRTV is smaller, making it closer to the actual HDRTV. Additionally, the reduced FHAD and NHQE metrics suggest that the HDRTV produced by our method better matches the distribution of HDRTV captured from real-world scenes.

Furthermore, we performed a qualitative comparison in Fig.\ref{figinsyn}. Previous methods failed to produce satisfactory conversion results, such as HDRTVDM and FMNet. Our method can convert HDRTV with more realistic and natural highlights and is particularly effective in reinstating natural and continuous color gradients in areas experiencing color transitions. 
This capability ensures that the images not only exhibit greater visual fidelity but also reflect a smoother and more authentic representation of real-world colors and shading, enhancing the overall viewing experience.

\subsubsection{Visual Perceptual Quality Assessment via HDRVDP3 metrics}

Table \ref{tabhdrvdp3} presents a comparative performance analysis of various methods on the HDRTV1K, HDRTV4K, and SRITM datasets using the HDRVDP3 metric.

Our method achieves the highest HDRVDP3 scores on the HDRTV1K (8.28), HDRTV4K (7.91)  and SRITM(7.8) datasets, indicating superior perceived quality of the HDRTV content.

\subsubsection{Experimental Results on Real Datasets}
To verify the generalization of our method in the real world, we evaluate it on multiple real-world datasets, including BSD100 \cite{937655}, CBSD68-Noisy \cite{937655}, CBSD68-Original \cite{937655}, Set5 \cite{set5bi}, and Set14 \cite{set14bi}.
As shown in Table \ref{tabrealtesttable}, our RealHDRTVNet achieves the best FHAD score as well as NHQE perceptual quality on real-world SDR datasets.

\begin{table}[htbp]
  \caption{Ablation study results for feature modulation encoder $E_{sfm}$, HDR Color Alignment module $HCA$ and the SDR Texture Alignment module $STA$.}
  \centering
    \begin{tabular}{cccccc}
    \toprule
    $E_{sfm}$  & $HCA$   & $STA$ & PSNR$\uparrow$ & LPHPS $\downarrow$ & FHAD $\downarrow$ \\
    \midrule
    $\checkmark$        &                   &              & 34.92 & 0.1186  & 91.62\\
    $\checkmark$        & $\checkmark$      &              & 34.95 & 0.1183  & 91.58\\
                        & $\checkmark$      & $\checkmark$ & 35.01 & 0.1173  & 91.53\\
    $\checkmark$        & $\checkmark$      & $\checkmark$ & 35.06 & 0.1166  & 91.03\\
    \bottomrule
    \end{tabular}%

  \label{tababl1}%
\end{table}%

\subsection{Ablation Study}

\begin{figure}
    \centering
    \includegraphics[width=0.9\linewidth]{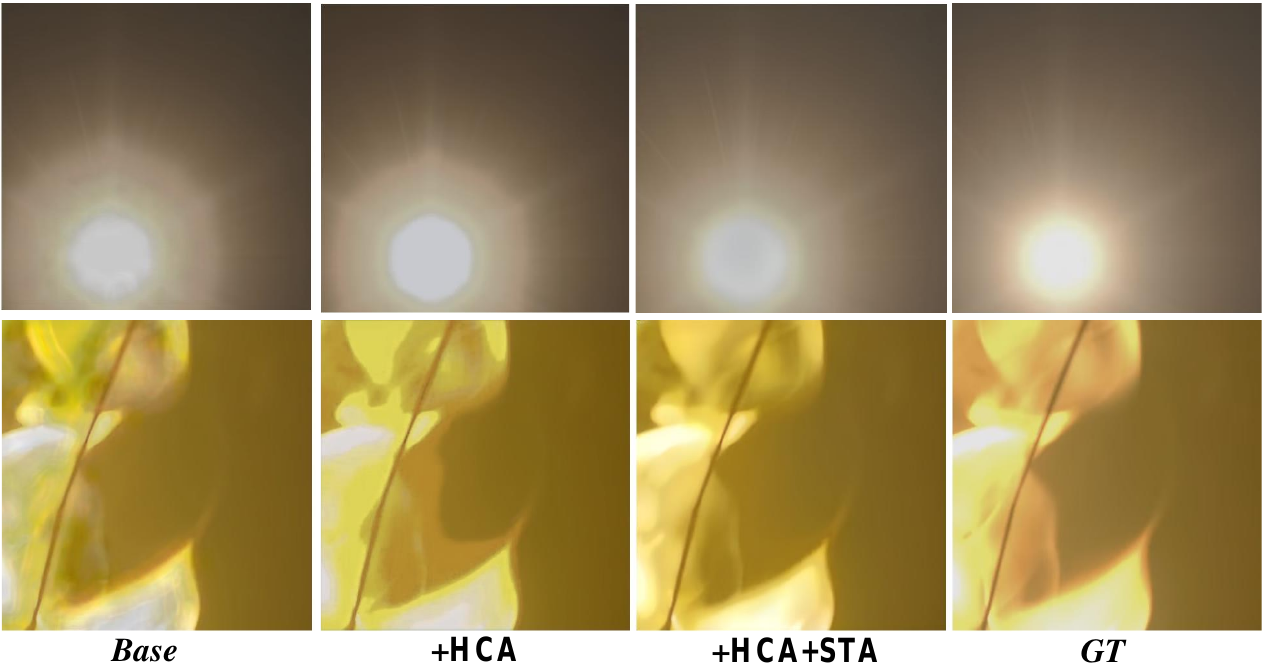}
    \caption{Visual ablation. The HDR Color Alignment $HCA$ module and the SDR Texture Alignment $STA$ module are added.}
    \label{fig22222}
\end{figure}

To assess the impact of the proposed components, we start with a baseline and systematically integrate each component one by one. In Table \ref{tababl1}, we can observe that each component has brought improvement, and the improvement in the PSNR metric reached 0.14. Meanwhile, our visual ablation results, presented in Fig. \ref{fig22222}, demonstrate that the modules we designed significantly enhance visual quality.

\subsubsection{Importance of SDRTV Feature Modulation Encoder $E_{sfm}$}
We first study the effectiveness of feature modulation encoders. As shown in the third line of Table \ref{tababl1}, deleting the modulation module in the encoder will cause the LPIPS and PSNR metric to deteriorate.

\subsubsection{Effectiveness of HDR Color Alignment $HCA$ and SDR Texture Alignment $STA$}
We ablate the brightness prior extraction module $HCA$ and the SDR Texture Alignment module $STA$.
As can be shown in Table \ref{tababl1}, after adding $HCA$ and module $STA$ in sequence, the performance of our model improved.

\section{Conclusion}

In this work, we introduce a novel paradigm for SDRTV-to-HDRTV conversion: HDRTV prior-guided high-quality SDRTV-to-HDRTV transformation. In contrast to traditional approaches that solely rely on SDRTV, our method achieves a more realistic and superior quality in HDRTV reconstruction. Additionally, we extend commonly used subjective quality evaluation metrics in SDRTV, such as FHAD, NHQE, and LPHPS, to assess the quality of HDRTV. Our proposed technique exhibits significant improvements in visual quality.

{
    \small
    \bibliographystyle{ieeenat_fullname}
    \bibliography{main}
}

\end{document}